\documentclass[cits]{PoS}

\title{Long-term AGN variability and the case of GSN~069}

\ShortTitle{Long-term AGN variability}

\author{\speaker{Richard Saxton}\\
        XMM-Newton SOC, ESAC, Apartado 78, E-28691 Villanueva de la Ca\~nada, Madrid, Spain\\
        E-mail: \email{richard.saxton@sciops.esa.int}}

\author{Andrew Read\\
        Dept. of Physics and Astronomy, University of Leicester, Leicester LE1 7RH, U.K.\\
        E-mail: \email{amr30@star.le.ac.uk}}

\author{Pilar Esquej\\
        Centro de Astrobiolog\'{i}a (CSIC-INTA), E-28850 Torrejon de Ard\'{o}z, Madrid, Spain\\
        E-mail: \email{pilar.esquej@cab.inta-csic.es}}

\author{Giovanni Miniutti\\
        Centro de Astrobiolog\'{i}a (CSIC-INTA), ESAC, Apartado 78, E-28691 Villanueva de la Ca\~nada, Madrid, Spain\\
        E-mail: \email{gminiutti@cab.inta-csic.es}}

\author{Emilio Alvarez\thanks{Current address: Universidad Complutense, E-28040 Madrid, Spain}\\
        ESAC, Apartado 78, E-28691 Villanueva de la Ca\~nada, Madrid, Spain\\
        E-mail: \email{ealvarez@gae.ucm.es}}

\abstract{We report on a study of long-term flux variations in a sample 
of more than 1000 AGN observed with ROSAT and in the XMM-Newton 
slew survey. Over a period of 3-19 years, NLS1 galaxies as a class are found to 
be only slightly more variable than broad line Seyfert galaxies,
despite the strong short term variability seen in some bright nearby NLS1s. 
Contrary to expectations, it is Seyfert II galaxies that exhibit the 
greatest flux volatility. 
One particular Sy II, which has brightened 
by a factor $>300$ over 15 years, has been monitored in detail
with Swift and XMM. The spectrum is extremely soft (kT $\sim60$eV) consistent 
with pure thermal emission from an accretion disk and reminiscent of 
ROSAT observations of the 
NLS1, WPVS 007. We show that this is likely to be a "true" Sy II, without
a BLR, and speculate that in its new high luminosity state we may be able
to witness a BLR in formation.} 

\FullConference{Narrow-Line Seyfert 1 Galaxies and their place in the Universe - NLS1,\\
		April 04-06, 2011\\
		Milan Italy}

\begin{document}

\section{Introduction}
The variability of the radiation output of AGN is well established and is
one of their defining characteristics. In single observations of
Seyfert I galaxies, fluctuations in the X-ray output on timescales of minutes
to hours is the norm (e.g. NGC~4051 \cite{Vaughan11} and MRK~766 \cite{Markowitz07}). The power spectral density function
(PSD) of
these variations can be described by red noise;
a smooth spectrum with increasing amplitude at lower frequencies
(\cite{McHardy87}) .
The PSD of AGN has been shown to have a break, where the slope flattens, 
 occurring on a timescale that scales with
the luminosity and by implication the black hole mass. 
Lower mass (luminosity) AGN
have a break in the PSD of $\sim10^{-2}$ days (e.g. NGC 4051 \cite{McHardy04}) 
whereas the most massive AGN have breaks at $\sim100$ days (e.g.
NGC 3516 \cite{Edelson99}).
This means that on short timescales, low mass AGN exhibit
stronger X-ray variability than their more massive counterparts
whereas on timescales of years the variability should be consistent.
An additional dependence on the fraction of the Eddington accretion rate,
$\dot{m}$ has been observed, leading to the
refinement that the variability can be explained by
characteristic pertubations in a steady-state accretion disk, whose
inner radius scales with $M_{BH} / \dot{m}$ \cite{McHardy06}.

The class of objects known as NLS1, show stronger variability than Sy I
and exhibit more rapid variations. This is consistent with the belief that 
they are low mass systems accreting at a high fraction of the Eddington rate.

Very large changes in flux due to variations in line-of-sight absorbing
material have been observed in several sources. In the Seyfert 1.8
galaxy, NGC1365, factor 10
variations, accompanied by strong spectral changes have been seen several times
(\cite{Risaliti05}, \cite{Risaliti09}) and can be modelled by movements of individual clouds in the broad-line region.

Finally, Bl Lacs or Blazars have been famous for their volatility at
all wavelengths for a long time (e.g. \cite{Giommi}). 

While there has been a lot of work on short-timescale variations in specific
sources and longer baseline studies of bright sources with RXTE, to date
there have been few systematic studies of the variability of
large samples of objects. This is mainly due to the lack of large-area
sensitive sky surveys. The ROSAT All Sky Survey (RASS \cite{voges}) is an excellent
resource for AGN with $F_{0.2-2}>3\times10^{-13}$ erg s$^{-1}$ cm$^{-2}$
but it can only be systematically
compared with data taken from ROSAT pointed observations taken months
or a few years after the RASS completed in 1990. With the advent of the
XMM-Newton slew survey (XMMSL1 \cite{Saxton08}) containing observations
from 40\% of the sky with $F_{0.2-2}>6\times10^{-13}$ erg s$^{-1}$ cm$^{-2}$,
a sensible comparison of AGN
flux over a baseline of 3-19 years can now be achieved over 40\% of the sky.
In this paper we make a systematic comparison of the long-term variability
of all classes of AGN detected either in the ROSAT or XMMSL1 surveys.
A $\lambda$CDM cosmology with ($\omega_{M},\omega_{\lambda}$) = (0.3,0.7)
and  $H_{0}$=70 km$^{-1}$sec$^{-1}$ Mpc$^{-1}$ has been assumed throughout.

\section{Method of ROSAT/XMM comparison}

We have made a sample of AGN which have been observed (but not necessarily detected)
in both the RASS, or ROSAT pointed observations, and XMMSL1. 
We define three groups of objects:

\begin{itemize}
\item XMMSL1 sources, identified as AGN or galaxies, which have counterparts
in ROSAT.
\item XMMSL1 sources, identified as AGN or galaxies, which have been observed
but not detected by ROSAT.
\item RASS sources, identified as AGN or galaxies, which have been observed
but not detected in XMMSL1.
\end{itemize}

We have used sources from the XMMSL1-delta3 CLEAN catalogue, released in August 2009, containing slew data taken between August 2001 and January 2009.

Sources which are detected with extended X-ray emission in either XMM-Newton
or ROSAT have been excluded.

\subsection{Upper limits}

For the XMMSL1 AGN without a ROSAT counterpart a 2-sigma upper limit to the
ROSAT count rate has been found using the {\tt EXSAS} software package
(\cite{Zimmerman}).

For RASS AGN, observed but not detected in XMMSL1, we have calculated a
2-sigma upper limit to the XMM-Newton count rate using a web-based upper limit server\footnote{
http://xmm.esac.esa.int/external/xmm\_products/slew\_survey/upper\_limit/uls.shtml} (\cite{SaxUpper}).

The final sample contains 1038 AGN of which 689 were detected in both surveys,
223 have a RASS upper limit and 126 have an XMMSL1 upper limit.

\subsection{Count rate to flux conversion}
\begin{figure}
\centering
\rotatebox{-90}{\includegraphics[height=7cm]{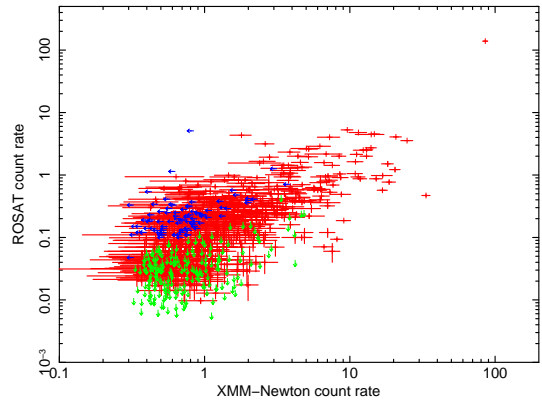}}
\caption[A comparison of the ROSAT and XMM-Newton slew count rates]
{ \label{fig:roscnts} A comparison of the ROSAT and XMM-Newton (0.2--2 keV) 
slew count rates}

\end{figure}

\begin{figure}
\centering
\rotatebox{-90}{\includegraphics[height=7cm]{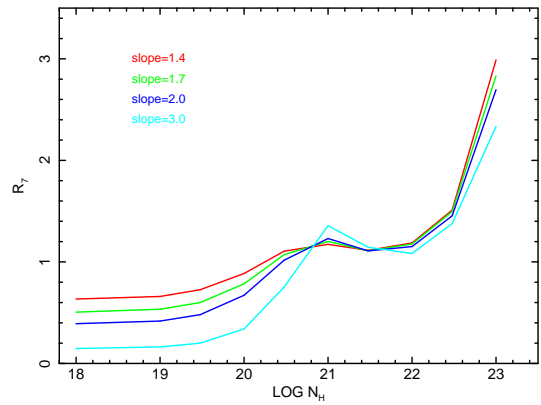}}
\caption[The influence of source spectra (an absorbed power-law) on the 
XMMSL1 / ROSAT count rate ratio]
{ \label{fig:specfac} The influence of source spectra (an absorbed power-law)
 on the XMMSL1 / ROSAT count rate ratio}

\end{figure}

\begin{figure}
\centering
\rotatebox{0}{\includegraphics[height=6cm]{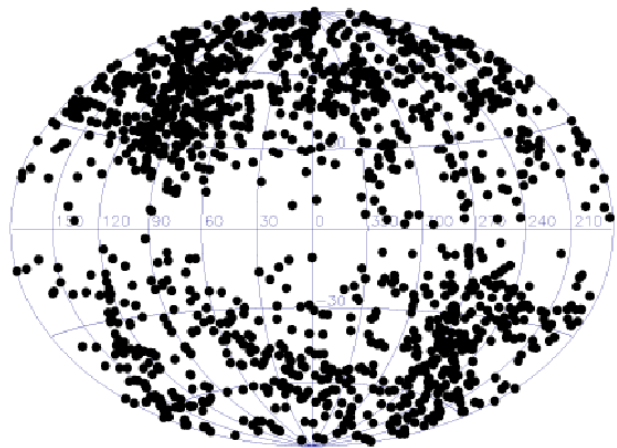}}
\caption[The sky distribution of the AGN in the sample in Galactic coordinates]
{ \label{fig:skydist} The sky distribution of the AGN in the sample in Galactic coordinates}

\end{figure}

All XMM-Newton slew observations have been made with the EPIC-pn detector
using the Medium filter.
In Figure~\ref{fig:roscnts} we plot the XMMSL1, 0.2--2 keV count rate
against the RASS count rates.
We find the mean ratio of the XMM to RASS count rates is 6.94+/-0.20.
From now on we use a factor 7 to normalise the count rate ratio, such that an intrinsic ratio of 7 is considered to show the same 0.2-2 keV flux in the two missions. The conversion factor from count rate to flux is crucial for comparing fluxes from different instruments and we need to be very careful when considering the effect of the source spectrum on this conversion.
In Figure~\ref{fig:specfac} we show the count rate ratio (normalised by the factor 7; $R_{7}$) as a function of the source spectrum. In principle there is a 
large variation of $R_{7}$ if the absorbing column is very 
low or very high. 
With these AGN observations, NH is practically limited to a minimum 
of $\sim 1\times10^{20}$ cm$^{-2}$, due to our Galaxy, and a maximum of  
$\sim10^{22}$ cm$^{-2}$ due to the flux limitations of the two surveys\footnote{
This applies to cold absorption. The behaviour of ionized absorbers which 
permit the passage of low-energy X-rays is quite different}. 
This can be seen clearly in the sky distribution of the objects in the
sample where few objects lie in the Galactic plane (Fig.~\ref{fig:skydist}). 
Similarly we
can infer that the intrinsic absorption can not much exceed 
$\sim10^{22}$cm$^{-2}$.
In this range, and for spectral slopes of 1.0 to 3.0, we see that a source 
with equal flux in RASS and XMMSL1 observations will have $0.3<R_{7}<$1.5.

We now define the variability ratio, $R_{V}$, to be the ratio 
between the higher and lower count rate irrespective 
of the observatory\footnote{Hence $R_{V}=1$ when $R_{7}=1$ and $R_{V}=2$ when $R_{7}=2$ or $R_{7}=0.5$.}

\section{Results of ROSAT/XMM comparison}

From the full population of 1038 AGN we find a median count rate ratio 
($\hat{R_{V}}$) of
2.11, with 68.4\% of the sources having $R_{V} \leq 3$ and 4.9\%
with $R_{V}\geq 10$.

We have made simulations of a non-varying population of sources with
the Galactic $N_{H}$ randomly selected from the actual sky values of the
1038 AGN, 
power-law indices selected from the population of ROSAT AGN slopes reported 
in \cite{Mittaz99} ($\Gamma=2.05\pm0.55$)
and count rates randomised using the actual errors in the XMMSL1 
and RASS measurements. The simulated sources have a variability ratio 
with a median value $\hat{S_{V}}$=$1.295\pm{0.003}$.

Following, \cite{Almaini} and \cite{Mateos07} we calculate
the true variability signal ($V$) by:

\begin{equation}
   V^{2} = (R_{V}-1)^{2}  -  (S_{V}-1)^{2}
\end{equation}

where $S_{V}$ is effectively the noise due to the
spectral effects and measurement errors.
This gives a median variability, $V$, of  $107\pm{7}$\%
in the 0.2--2 keV energy band. In the UV band the mean long-term variability
was found to be 35\% from a survey of 9000 SDSS
quasars (\cite{McCleod10}). This result then continues the trend
of higher photon 
frequencies exhibiting larger amplitude variability, seen on shorter timescales
 (e.g. NGC 5548; \cite{Clavel}).

\subsection{Luminosity dependence}

\begin{figure*}
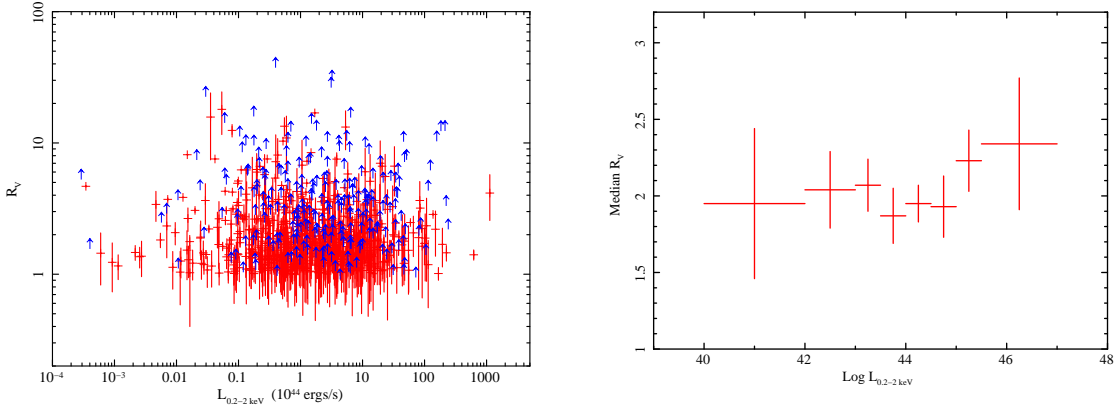

\begin{minipage}{3in}
    \rotatebox{-90}{\includegraphics[width=5.5cm]{lum_ratio_plot.ps}}
\end{minipage} \hspace*{0.1in}
\begin{minipage}{3in}
    \rotatebox{-90}{\includegraphics[width=5cm]{lumbinned.ps}}
\end{minipage} 
\caption{ \label{fig:xlumvar} The X-ray variability of our sample as a
function of XMMSL1 0.2--2 keV luminosity; left panel: individual sources,
right panel: median variability per luminosity bin.}
\end{figure*}

In Fig.~\ref{fig:xlumvar} we plot $R_{V}$ against the X-ray luminosity for 
the sources where the redshift is known. No significant correlation is 
found. As we are probing variability timescales far longer than the PSD break timescale it is normal that the luminosity-dependence, seen at shorter timescale, has no influence. This result shows that no further luminosity (mass) 
dependent effect is affecting the longer-term variability.  

\subsection{Source category}

We have cross-correlated our sample with the Veron catalog of Quasars \& AGN
\cite{Veron} to find the AGN class for each source, where known.
In Table 1 we separate the sources by AGN class and calculate
the median variability and the fraction of sources showing $R_{V}<2$,
$<3$ and $>10$. The calculations ignore sources with upper limits that 
fall outside of these limits.


Seyfert II galaxies have the highest percentage of sources showing
strong ($R_{V}>10$) variability. It is not clear whether this is due 
to changes in line-of-sight absorption or changes in the intrinsic emission
and these data can not help distinguish between the two possibilities.
A closer look at one particular example is needed.

\begin{table}
\caption{Variability statistics for each AGN class}
\label{tab:stats}      
\begin{center}
\begin{tabular}{l c c c c c}
\hline\hline                 
CLASS$^{a}$ & No.$^{b}$ & Median$^{c}$ & \multicolumn{3}{c}{Fraction (\%)$^{d}$} \\
      &           &  Var. (\%) & $R_{V}<2$ & $R_{V}<3$ & $R_{V}>10$  \\
\\
ALL   & 1038      &   $107\pm{7}$ &  46.8  & 68.4 & $4.9\pm{0.8}$ \\
S1    &  318      &   $90\pm{12}$ & 51.8   &  72.0 & $4.1\pm{1.4}$ \\
S1.2  &   48      &   $66\pm{27}$ & 62.5   &  78.9 & 0.0 \\
S1.5  &   83      &   $112\pm{26}$ & 45.1   &  65.7 & $3.3\pm{2.4}$ \\
S1.8/1.9/2 &  37  &   $155\pm{44}$ & 34.3   &  55.9 & $13.3\pm{6.0}$ \\
QSO   &  232      &   $130\pm{18}$ & 42.4   &  62.1 & $4.8\pm{2.0}$ \\
NLS1  & 64        &   $99\pm{28}$ & 48.1   & 72.5   &   $4.1\pm{2.9}$ \\
Blazar& 142       &   $109\pm{19}$ & 43.9  & 73.4   &   $3.4\pm{1.7}$ \\
\hline                        
\end{tabular}
\\
\end{center}
$^{a}$ AGN class: Seyfert 1, 1.2, 1.5, a combination of 1.8, 1.9 and 2, QSO,
NLS1 and a combination of Blazars, blazar candidates and highly polarised
quasars.\\
$^{b}$ The total number of sources in this category. \\
$^{c}$ Median variability, $\hat{V}$, expressed as a percentage, 
after correcting for spectral effects (see text).\\
$^{d}$ The percentage of sources with $R_{V}$ less than 2, less than 3 or 
greater than 10. \\
\end{table}

\section{GSN~069}
\begin{figure}
\centering
\rotatebox{-90}{\includegraphics[height=7cm]{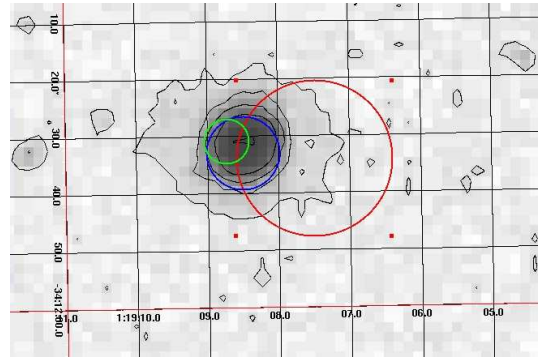}}
\caption[Image of GSN~069]
{ \label{fig:gsnim} A DSS image and contours of GSN~069: the red circle represents the XMM-Newton slew error circle, blue is for the SWIFT-XRT position and green is the SWIFT position enhanced using UVOT}
\end{figure}

During a slew on July 14, 2010 XMM-Newton detected soft X-ray emission 
at $1.5\pm{0.3}$ counts s$^{-1}$
from a position consistent with the nucleus of GSN~069 (also known 
as 6dFg0119087-341131; z=0.01816) (see Fig.~\ref{fig:gsnim}). 
Previous ROSAT survey and pointed observations
failed to detect the source and give upper limits, to the unabsorbed flux,
 30--360 times
fainter than the XMM-Newton slew detection (Table~\ref{tab:obslog}).

\begin{figure}
\centering
\rotatebox{0}{\includegraphics[height=6cm]{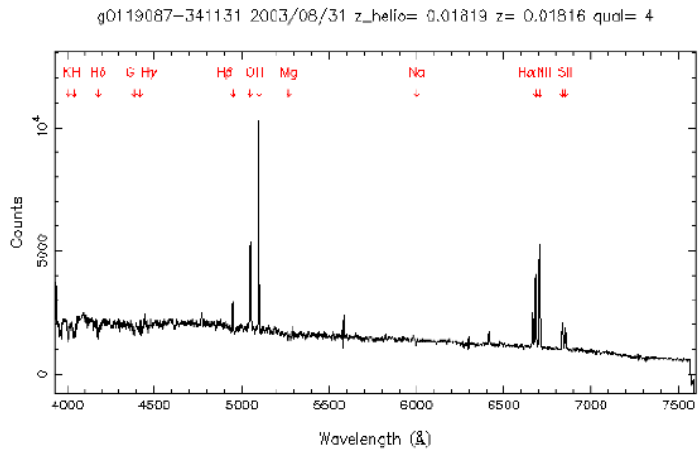}}
\caption[Optical spectrum of GSN~069]
{ \label{fig:gsnoptical} Optical spectrum of GSN~069 from the 6dF survey} 
\end{figure}

Two, consistent, optical spectra, taken in 2001 and 2003 in the
2dF and 6dF surveys \cite{6dF}
show unresolved Balmer lines ($FWHM\le200$ km s$^{-1}$)
and line ratios consistent with a Sy 2 classification (Fig.~\ref{fig:gsnoptical}).

\begin{figure}
\centering
\rotatebox{-90}{\includegraphics[height=7cm]{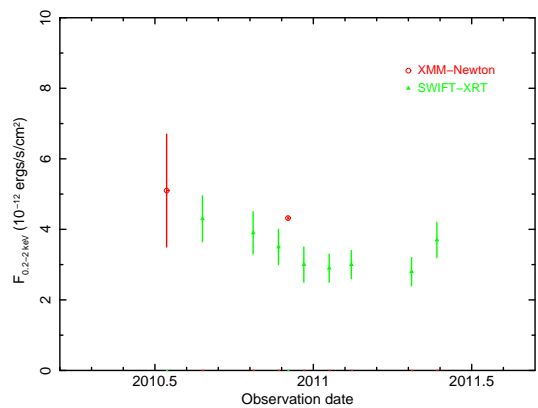}}
\caption[Light curve of GSN~069]
{ \label{fig:gsnlc_long} The X-ray light curve of GSN~069}

\end{figure}

The source has been monitored with the SWIFT-XRT since its discovery. 
The SWIFT observations have been analysed following the procedure outlined in 
Evans et al. (2009) \cite{evans} and
show stable soft X-ray emission to within a factor 2 (Fig.~\ref{fig:gsnlc_long}).

\begin{table}
\caption{GSN~069 observation log}
\label{tab:obslog}      
\begin{center}
\begin{tabular}{l c l l}
\hline\hline                 
Mission$^{a}$ & Date & Count rate$^{b}$ & Flux$^{c}$ \\
\\
RASS  & 1990    &   $<0.0099$ &  $<0.11$  \\
ROSAT-PSPC    &  1993-07-13T04:38:55  &   $<0.00177$ & $<0.019$   \\
ROSAT-PSPC    &  1994-06-29T05:59:51  &   $<0.00129$ & $<0.014$   \\
XMM slew  &   2010-07-14T00:48:14      &   $1.49\pm{0.35}$ & $5.1\pm{1.6}$   \\
SWIFT  & 2010-08-27T06:05:40        &   $0.041\pm{0.005}$ & $4.3\pm{0.6}$ \\
SWIFT  & 2010-10-27T05:32:30        &   $0.037\pm{0.004}$ & $3.9\pm{0.6}$ \\
SWIFT  & 2010-11-24T01:19:48  &   $0.033\pm{0.004}$ & $3.5\pm{0.5}$ \\
XMM pointed &   2010-12-02T10:44:18      &   $1.03\pm{0.01}$ & $4.32\pm{0.04}$\\
SWIFT  & 2010-12-22T15:05:03        &   $0.029\pm{0.004}$ & $3.0\pm{0.5}$\\
SWIFT  & 2011-01-19T06:15:29        &   $0.028\pm{0.004}$ & $2.9\pm{0.4}$\\
SWIFT  & 2011-02-16T05:33:31        &   $0.029\pm{0.003}$ & $3.0\pm{0.4}$ \\
SWIFT  & 2011-04-25T05:00:28        &   $0.027\pm{0.003}$ & $2.8\pm{0.4}$ \\
SWIFT  & 2011-05-23T04:06:01        &   $0.035\pm{0.004}$ & $3.7\pm{0.5}$ \\
\hline                        
\end{tabular}
\\
\end{center}
$^{a}$ XMM-Newton, EPIC-pn camera: slew observation performed in FullFrame 
mode with 
the Medium filter; pointed observation performed in FullFrame mode with 
the Thin filter. SWIFT-XRT in pc mode.\\
$^{b}$ counts s$^{-1}$ in the band 0.2--2 keV. \\
$^{c}$ Unabsorbed flux, $F_{0.2-2~keV}$, units of $10^{-12}$ erg s$^{-1}$ cm$^{-2}$, 
calculated from a black body model
with kT=58 eV and Galactic absorption ($2.48\times10^{20}$cm$^{-2}$) \\
\end{table}

\begin{figure}
\centering
\rotatebox{-90}{\includegraphics[height=7cm]{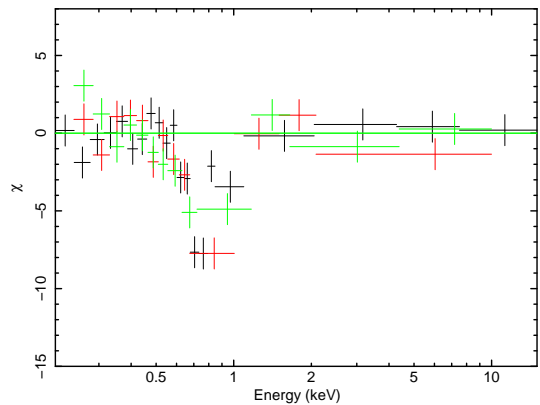}}
\caption[GSN 069 - spectral edge]
{ \label{fig:gsnEdge} Residuals to a {\em diskbb} model fit to the 
XMM-Newton EPIC-pn (black), MOS-1 (red) and MOS-2 (green) 
data} 

\end{figure}

\subsection{X-ray spectral fits}
An XMM-Newton TOO, with nominal exposure 10ks, was performed in December 2010,
using all of the X-ray detectors and the optical monitor. 
The data were reduced using XMM-SAS v11.0 (\cite{Gabriel}). EPIC source
spectra were extracted from a circle of 30 arcsecond radius and 
background taken from a nearby source-free region. 
The two RGS cameras were combined into one
spectrum. All the camera spectra were grouped so as to contain 
a minimum of 25 counts per bin  and to oversample the instrument 
resolution by a maximum of 3.
Fits to the individual instruments gave consistent results and
so all spectra were fit simultaneously, with a constant used to compensate
for the small normalisation differences.

An initial fit with a power-law, absorbed by the Galactic column ($2.48\times10^{20}$; LAB Map - \cite{Kalberla}) gave a poor fit but showed that the
overall spectrum is very soft ($\Gamma=6.2$).
To model the shape better we fit the soft X-rays with a black-body or
multi-colour disk model (diskbb in XSPEC) and the weak 
emission at $>2$ keV with a power-law (with $\Gamma$ fixed to 1.7).
The black-body and diskbb models yielded better fits 
with $\chi_{r}^{2}=1.4, 2.0$ respectively (Tab.~\ref{tab:gsnfits}). 
To fit the remaining residuals we tried various absorption models.
Cold absorption, in the rest frame of the source, is excluded with an 
upper limit of $9\times10^{19}$cm$^{-2}$.
Good fits were obtained with a simple edge of energy $650$ eV or with 
an ionized absorber with $N_{H}$=4-11$\times10^{23}$cm$^{-2}$,
$\xi=100-800$ and a covering fraction of 74--100\% redshifted by 
either 0.15 or 0.2 in the rest frame of the source, depending on 
the continuum model
(Fig.~\ref{fig:gsnEdge}).

The unabsorbed soft X-ray luminosity is $L_{0.2-2}=5.3\pm{0.1}\times10^{42}$ 
erg s$^{-1}$
for the best fit model (diskbb with edge). This model greatly underpredicts
the flux observed in the OM UVW1, UVM2 and B filters. However, the optical
images of the galaxy appear extended in all filters and so the count 
rates are clearly
dominated by emission from the stellar population. We
therefore extrapolate the soft X-ray spectral model over optical, UV and EUV 
frequencies to obtain $L_{bol}=1.8\pm{0.1}\times10^{43}$ erg s$^{-1}$.
The emission above 2 keV is very weak, $1.1\pm{0.9}\times10^{-3}$ counts s$^{-1}$ in the EPIC-pn detector, and the conversion of this count rate to an unabsorbed flux is 
highly model 
dependent. We find a highest, 90\% confidence, upper limit of 
$L_{2-10}<8.5\times10^{40}$ erg s$^{-1}$,  for the black-body with partial 
covering model.

\begin{table}
\caption{GSN~069 spectral fits to an XMM-Newton TOO}
\label{tab:gsnfits}      
\begin{center}
\begin{small}
\begin{tabular}{c c c c | c c c c c c |l}
\hline\hline                 
 \multicolumn{4}{c}{Low-energy model} & \multicolumn{6}{c}{Intrinsic absorption} & $\chi^{2}$/dof  \\
\\
Plaw  &  Bbody  &  \multicolumn{2}{c}{Diskbb}  &  \multicolumn{2}{c}{Edge$^{a}$}  & \multicolumn{4}{c}{zxipcf$^{b}$}  \\
$\Gamma$   &  kT (ev)      &  kT (eV) & Norm & $E$ (eV) & Tau & $N_{H}$ & $\xi$ & cf(\%) & z \\
\hline\hline                 
6.2   & -         &  -      &  -   &    -   &  -  &  - & - & - & - & 2015/107  \\
-   & $56.3^{+0.7}_{-0.3}$ &  -      &  -   & -  &   -  &  - & - & - & - & 152/107 \\
-   & $58.2^{+0.7}_{-0.5}$ &  -      &  -   &  651$^{+22}_{-13}$        &  $0.80^{+0.27}_{-0.19}$  &  
- & - & - & - & 116/105 \\
-   & $62.2^{+0.7}_{-0.6}$ &  -      &  -   & -  &  - & $98^{+10}_{-18}$ & $2.7^{+0.1}_{-0.7}$ & $82^{+4}_{-8}$ &
 $0.167^{+0.005}_{-0.008}$ & 114/103 \\
-   & -                    & 67.1      &  43695   &  -        &   -  &  - & - & - & - & 215/107 \\
-   & -                    & $71.7^{+0.9}_{-0.9}$  & $28184^{+370}_{-470}$   & 650$^{+16}_{-11}$  & $1.44^{+0.29}_{-0.24}$    &  - & - & - & - & 112/105 \\
-   & -      & $77.4^{+1.1}_{-2.0}$  & $21105^{+340}_{-460}$   & - & - & $54^{+9}_{-10}$  & $2.8^{+0.1}_{-0.6}$ & $92^{+8}_{-8}$ & $0.225^{+0.007}_{-0.007}$  & 116/103 \\
\hline                        
\end{tabular}
\\
\end{small}
\end{center}
All fits included absorption by the Galactic column ($N_{H}=2.48\times10^{20}$cm$^{-2}$) and a power-law with slope fixed at $\Gamma=1.7$ to model the very weak 
emission beyond 2 keV. Errors are 90\% confidence. \\
$^{a}$ The {\em edge} model in {\em XSPEC}. 
$^{b}$ A partially ionized, partial covering model ({\em zxipcf} in {\em XSPEC}) with parameters, $N_{H}$ in
units of $10^{22}$cm$^{-2}$, log of the ionization parameter, covering 
fraction and redshift.\\ 
\end{table}

A spectral fit of the combined SWIFT-XRT data gives very similar spectral 
parameters to the XMM-Newton spectrum showing that the soft spectral
shape of GSN~069 is long lasting. 

A possible explanation for the steep X-ray spectra of GSN~069 is 
provided by an ionized
absorber which allows very soft X-rays to pass while strongly blocking
higher energies. This model was used to explain the very steep 
spectra of WPVS~007 seen during the ROSAT era \cite{Grupe}.
To test for this we have attempted to fit the GSN~069 XMM-Newton spectrum 
with 
intrinsic power-law emission absorbed by up to three ionized, partial 
covering, absorbers. No good fit was obtained with a power-law 
model with $\Gamma<4$ and therefore we conclude that the very steep soft X-ray spectrum is thermal in origin.

\subsection{Discussion}

The lack of a cold absorber in the X-ray spectrum of GSN~069 is difficult 
to reconcile with the classical interpretation of a Sy II where the broad 
lines are removed by absorption. From the 6dF optical spectrum we find 
a ratio of [OIII]5007 / H$_{\beta} \sim11$ which is symptomatic of
a missing BLR rather than absorption in Sy II \cite{Hawkins}. 
From this we conclude that the BLR
is absent rather than obscured in this case.
In other words, GSN 069 appears to be another example of an AGN
classified as a Sy II galaxy but with little or no intrinsic
absorption in the X-rays and with optical line ratios inconsistent
with heavy extinction. The
X-ray spectrum could be reconciled with a Sy II classification if the
AGN was Compton-thick in the X-rays (although the optical line ratios
do not support this idea). However the soft X-ray variability on both
short (doubling time of 800s in the XMM-Newton observation) and long
(see Table 2) timescales rules
out this hypothesis, making GSN 069 a ``true Sy II'' candidate, i.e. a
candidate AGN with no BLR.

In the paradigm where the BLR is created from outflows of material from
the disk, the outflow may not be sustainable and hence the BLR may be absent 
if $L_{bol}<10^{42}$ erg s$^{-1}$ \cite{Elitzur06} or if the accretion 
rate is below a critical value $\dot{m}<0.01$ \cite{Nicastro}.

The black hole mass, $M_{BH}$, can be estimated from the normalisation 
of the {\em diskbb} model if the emission is thermal, as it seems to be.
Following Yuan et al. 2010 \cite{Yuan}, and taking the disk inclination 
as i=40$^{\circ}$ for simplicity, we 
derive the black hole mass as
$M_{BH}=1.7\times10^{5}$ M$_{\odot}$ for a non-rotating black 
hole and $M_{BH}=8.4\times10^{5}$ M$_{\odot}$ for a maximally 
spinning Kerr black hole. 
The accretion rate may be derived from the z-corrected temperature, 
to be $\dot{m}=0.017-0.38$ depending on the black hole spin.
An independent estimate of $M_{BH}$ may be found from its relationship 
with the bulge K band luminosity \cite{MarconiHunt}. The 2MASS
extended catalogue gives m$_{K}$=12.75, for the whole galaxy, which implies
$M_{BH}<2.5\times10^{6}$ M$_{\odot}$. 
The low mass of the black hole is also supported, to some extent, by fast
variability (doubling time of 800s) seen in the XMM-Newton light curve.

From the relationship between $M_{BH}$, $\dot{m}$ and the FWHM of the BLR
lines \cite{McHardy06} we would expect to be seeing broad lines with
FWHM=560-1800 km s$^{-1}$ in the optical spectrum. 

The luminosity of the [OIII]5007 line in the 6dF spectrum is 
$1.1\times10^{40}$ erg s$^{-1}$ which implies a historical $L_{bol}\sim10^{42}$ 
erg s$^{-1}$ using a bolometric correction factor of 88--142 \cite{Lamastra09};
an order of magnitude below the current value.
If radiating at an efficiency $\eta=0.1$, the historical accretion rate
will have been $\dot{m}=0.009-0.047$.
In the past then, the source has been accreting and radiating close to
the critical values necessary to produce outflows and plausibly
has been too weak to produce the BLR.
With the current AGN parameters an outflow should be sustainable, in which
case we can estimate the time necessary to form the 
BLR as $t=v/d$, where $v$ is the velocity of expanding material and $d$ is the
distance of the BLR which will be $\sim 10$ days for $L_{bol}=10^{43}$ erg s$^{-1}$ 
 \cite{Denny}. Hence for outflow velocities of 1000-6000 km s$^{-1}$ (e.g. \cite{Leighly})
BLR material will start to build up after 2-12 months. 
A regular monitoring
of the optical spectrum of GSN~069 may be able to detect this process.

The apparently redshifted, strong edge observed at $\sim650$ eV may be 
produced by a failed disk outflow or from an inflowing accretion disk
overlapping our line of sight. 

In summary, the soft emission from this SY II appears to be thermal in origin
and its variability appears to be intrinsic to the central engine and not
related to changes in line-of-sight absorption. This mechanism may also
be the cause of the strong variability seen in the other Sy II galaxies,
three of which have $L_{X}<10^{43}$ erg s$^{-1}$ in their high state.

\section{Acknowledgements}
The XMM-Newton project is an ESA science mission with instruments and contributions directly funded by ESA member states and the USA (NASA).
This work made use of data supplied by the UK Swift Science Data Centre at the
University of Leicester. We thank Jane Turner and Lance Miller for useful 
discussions about ionized absorber models.

\end{document}